\documentclass[aps,prl,twocolumn,groupedaddress]{revtex4}
\usepackage{graphicx}
\usepackage{amsmath,amsfonts,amssymb}
\DeclareMathOperator{\erf}{erf}
\usepackage{color}
\begin{document}
\title{Quantum Black Hole Wave Packet: \\
Average Area Entropy and Temperature Dependent Width}

\author{Aharon Davidson}
\email{davidson@bgu.ac.il}
\homepage[Homepage: ]{http://www.bgu.ac.il/~davidson}
\author{Ben Yellin}
\email{yellinb@bgu.ac.il}

\affiliation{Physics Department, Ben-Gurion University
of the Negev, Beer-Sheva 84105, Israel}

\date{April 14, 2014}

\begin{abstract}
A quantum Schwarzschild black hole is described,
at the mini super spacetime level,
by a non-singular wave packet composed of plane
wave eigenstates of the momentum Dirac-conjugate
to the mass operator.
The entropy of the mass spectrum acquires then
independent contributions from the average mass
and the width.
Hence, Bekenstein's area entropy is formulated using
the $\langle \text{mass}^2 \rangle$ average, leaving
the $\langle \text{mass} \rangle$ average to set the
Hawking temperature.
The width function peaks at the Planck scale for an
elementary (zero entropy, zero free energy) micro black
hole of finite rms size, and decreases Doppler-like towards
the classical limit.
\end{abstract}


\maketitle

Bekenstein-Hawking black hole area entropy \cite{BH}
constitutes a triple point in the phase of physical theories,
connecting gravity, quantum mechanics, and statistical
mechanics.
However, despite of several illuminating derivations
\cite{entropy}, the statistical roots of black hole entropy
have not been fully revealed, not even at the level of
discrete models \cite{BM}.
There exist a few extreme black hole solutions \cite{SV},
notably beyond general relativity, where one can apparently
count micro states.
But as far as the prototype Schwarzschild black hole is
concerned, we still do not have the finest idea where
these micro states are hiding, and how to enumerate
them.
A classical black hole is characterized by its event
horizon, but once $\hbar$ is switched on (to allow for
a finite Hawking temperature and non-zero Bekenstein
entropy), even the innocent looking question 'where is
this horizon located' lacks a decisive answer in the
quantum or even in the semi-classical level.

The quantum mechanical Schwarzschild black hole is
hereby described by a non-singular minimal uncertainty
wave packet composed of plane wave eigenstates.
We carry out our analysis at the mini super spacetime
level \cite{mini} without relying on theories beyond
general relativity such as string theory \cite{string},
the fuzzball proposal \cite{fuzzball}, or loop
quantum gravity \cite{loop}
(see \cite{Frolov} for a different approach).
Treating the black hole as a sub-system (a field theory
defined on a black hole background is expected to be
in a thermal state), its Gaussian mass spectrum becomes
temperature dependent.
We invoke Fowler prescription \cite{Fowler} for dealing
with such sub-systems, and show that the associated
statistical entropy acquires independent contributions
from the average mass as well as from the width, and
consistently formulate Bekenstein's area ansatz by
means of the $\langle mass^2\rangle$ average.
While, as expected, the $\langle mass\rangle$ average
turns out to be inversely proportional to Hawking
temperature, a novel temperature dependent width
function makes its appearance.
The width function is maximal at the reduced Planck
mass for an elementary quantum mechanical black hole
of finite rms size, for which both the entropy and free
energy vanish and are minimal, and decreases Doppler-like
towards the classical limit.

Let our starting point be the most general static radially
symmetric line element, expressed in the form
\begin{equation}
	ds^2=-\frac{y(r)}{2r}dt^2+\frac{2r}{x(r)}dr^2
	+r^2(d\theta^2+\sin^2 \theta d\phi^2) ~.
	\label{metric}
\end{equation}
The unfamiliar $x,y$ representation has been
carefully designed to avoid the appearance of explicit
$r$-dependence in the constrained Hamiltonian formalism
(see ref.\cite{GRG} for the canonically transformed
$r$-dependent Berry-Keating type \cite{Berry}
Hamiltonian).
A tenable gauge pre-fixing option, namely defining a
radial marker $r$ whose geometrical interpretation 
is $x,y$-independent, has been harmlessly exercised.
This has to be contrasted with the forbidden gauge
prefixing of the 'lapse' function (the coefficient of $dr^2$
in this case), which kills the Hamiltonian constraint and
introduces an unphysical degree of freedom
(no gauge pre-fixing in Kuchar's midi superspace approach
\cite{Kuchar}).
The more so at the mini super-spacetime level, where the
general relativistic action $\int {\cal R}\sqrt{-g}~d^4\,x$
is integrated out over time and solid angle into the mini
action $\int {\cal L}(x,x^{\prime},y,y^{\prime},r)dr$.

A word of caution is in order:
Throughout this paper we treat
$\int{\cal L}(q,q^{\prime},r)dr$ in full mathematical
analogy with $\int{\cal L}(q,\dot{q},t)dt$.
Technically, the $t$-evolution is traded for the
$r$-evolution, both classically as well as quantum
mechanically, with the notions of Lagrangian and
Hamiltonian being adapted accordingly.
A similar technique has been introduced by York and
Schmekel \cite{York}.
To sharpen the point, we clarify that our 'Hamiltonian'
(to be identified with the momentum Dirac-conjugate
\cite{Dirac} to the mass operator) has nothing to do with
the physical mass of the black hole.

Up to a total derivative and an overall absorbable factor,
the mini super-spacetime Lagrangian takes the form
\begin{equation}
	{\cal L}(x,x^{\prime},y,y^{\prime})=
	\left(\frac{3x^{\prime}}{4}
	-2\right)\sqrt{\frac{y}{x}}
	-\frac{y^{\prime}}{4}\sqrt{\frac{x}{y}} ~.
\end{equation}
Being linear in the 'velocities', it gives rise to two primary 
second class constraints, namely
\begin{equation}
	\phi_{y}= p_{y}+\frac{1}{4}
	\sqrt{\frac{x}{y}} \approx  0 ~, 
	~~ \phi_{x}= p_{x}-\frac{3}{4}
	\sqrt{\frac{y}{x}} \approx  0 ~,
\end{equation}
whose Poisson brackets do not vanish
$\displaystyle{\left\{\phi_{y}, \phi_{x}\right\}
=\frac{1}{2\sqrt{xy}}}$.
Following Dirac prescription \cite{Dirac}, we are
then driven from the naive Hamiltonian ${\cal H}
=p_x x^{\prime}+p_y y^{\prime}-{\cal L}
=2\sqrt{y/x}$ to the total Hamiltonian
\begin{equation}
	{\cal H}_{T}=2\sqrt{\frac{y}{x}}
	+2\frac{y}{x}\phi_{y}
	+2\phi_{x} ~.
\end{equation}
One can verify that the corresponding classical
solution is (and is nothing but) the Schwarzschild
solution
\begin{equation}
	\frac{y(r)}{2\omega^2 r}=\frac{x(r)}{2r}
	=1-\frac{2m}{r} ~,
	\label{Schwarzschild}
\end{equation}
with no restrictions on the sign of the integration
parameters $m$ and $\omega$.
Along the classical trajectories the Hamiltonian takes
the value ${\cal H}=2\omega$, telling us that the
${\cal H}$ is not the total physical mass of the system.

To quantize the system it becomes crucial to calculate
the Dirac brackets, and here one finds first of all
\begin{equation}
	\boxed{\vspace{10pt}\left\{x,y \right\}_D
	=2\sqrt{xy}\neq 0}
	\label{xy}
\end{equation}
Counter intuitively, and potentially with far reaching
consequences, two metric components do not Dirac
commute.
Moreover, the relation
$\left\{x ,\frac{1}{2}{\cal H}\right\}_D=1$
paves then the way for the quantum mechanical
commutation relations $[x,\frac{1}{2}{\cal H}]=i\hbar$.
${\cal H}$ is then faithfully represented by
\begin{equation}
	{\cal H}=-2i\hbar\frac{\partial}{\partial x}~.
\end{equation}
By the same token, in accord with eq.(\ref{xy}), the other
metric component $y$ is represented by
$y=\frac{1}{4}{\cal H}x{\cal H}$.

$\phi_{x,y}$ are second class constraints, so
$\phi_x \psi=\phi_y \psi=0$ are automatically fulfilled.
Denoting by $2\omega$ the eigenvalues of ${\cal H}$,
the corresponding eigenstates
are simple plane waves.
Their full $r$-'evolution' is given by
\begin{equation}
	\psi_{\omega}(x,r)=
	\frac{1}{\sqrt{4\pi}}e^{\frac{i}{\hbar}\omega (x-2r)}~.
	\label{plane}
\end{equation}
They are not localized and form a $\delta$-normalizable
set.
The most general solution is of the form $\psi(x-2r)$.
We are however after the 'most classical' wave packet
defined by the minimal uncertainty relation
$\Delta x \Delta {\cal H}=\hbar$, namely
\begin{equation}
	\psi(x,r)=
	\frac{e^{-\frac{(x-2r+4m)^2}{64\sigma^2}}}
	{2(2\pi)^{\frac{1}{4}}\sqrt{\sigma}}~,
	\label{psi}
\end{equation}
for which the classical Schwarzschild solution
eq.(\ref{Schwarzschild}) is both the average as well
as the most probable configuration.
We thus expect the wave packet eq.(\ref{psi})
to capture all semi-classical essence of black
hole thermodynamics.
We have limited ourselves in this paper to the 'most
classical' black hole wave packet simply because
Bekenstein-Hawking thermodynamics is formulated in
the background of a classical event horizon.
One can even construct an orthonormal tower of
non-minimal uncertainty wave packets \cite{GRG},
to be regarded a prediction of the mini
super-spacetime approach (to be discuss elsewhere),
none of which sharing the Schwarzschild configuration
as the most probable.
Eq.(\ref{psi}) is a superposition of plane
waves.
Its Fourier transform is given by
\begin{equation}
	\tilde{\psi}({\cal H})=
	\frac{2\sqrt{\sigma}}{(2\pi)^{\frac{1}{4}}}
	e^{-4\sigma^2 {\cal H}^2}e^{2im {\cal H}}~.
\end{equation}

We identify the mass operator as $M=\frac{1}{4}(2r-x)$
(in the ${\cal H}$-language it reads
$M=-\frac{1}{2}i\hbar\frac{\partial}{\partial {\cal H}}$).
For the Gaussian wave packet eq.(\ref{psi}) it means
\begin{equation}
	\langle M  \rangle=m~, \quad
	\langle M^2  \rangle=m^2+\sigma^2 ~.
\end{equation}
The black wave packet probability
density $\psi^{\dagger}\psi$ can be directly translated into
a statistical mechanics normalized mass spectrum
\begin{equation}
	\rho(M;m,\sigma)=
	\frac{e^{\frac{(M-m)^2}{2\sigma^2}}}{\sqrt{2\pi}\sigma}~.
	\label{rhoM}
\end{equation}
While a non-negative average mass $m\geq 0$ (the classical
choice) is soon to be dictated on thermodynamical grounds,
the mass distribution must cover, for the sake of quantum
completeness, the entire range $-\infty <M<\infty$.
However, the probability to have negative masses drops like
$\sim\exp(-m^2/2\sigma^2)$ towards the classical limit.

At this stage, one may wonder where is the black hole horizon
actually located?
As far as our wave packet is concerned, there is nothing
special going on in the neighbourhood of $r= 2Gm/c^2$
(and actually also not near the origin).
Supported by eq.({\ref{xy}), this suggests that a
sharp horizon is merely a classical gravitational concept.
However, one may still effectively
interpret eq.(\ref{rhoM}) as the quantum mechanical profile
of the horizon, with a probability density $\rho(M;m,\sigma)$
to find it at radius $2GM/c^2$.
In some sense, this reminds us of the fuzzball proposal
\cite{fuzzball} where the black hole arises from coarse
graining over horizon-free non-singular geometries.
See \cite{Casadio} for horizon wave packets,
and \cite{Marolf,York} for horizon fluctuations.

Treating the quantum mechanical black hole as a subsystem
(a field theory defined on a black hole background is expected
to be in a thermal state whose temperature at infinity is the
Hawking temperature), its Gaussian mass spectrum is
temperature dependent.
Following Fowler prescription for dealing with such a case,
the naive partition function must be modified according to
\begin{equation}
	Z(\beta)=\sum_n \rho_n e^{-\beta F_n (\beta)}~,
	\label{Fowler}
\end{equation}
with the Boltzmann factor being traded for the
Gibbs-Helmholtz  factor.
The Helmholtz free energy function $F$ obeys the
Gibbs-Helmholtz equation
\begin{equation}
	F+\beta\frac{\partial F}{\partial \beta}=E(\beta)
	~\Rightarrow~
	\beta F(\beta)=\int_{\beta_0}^{\beta} E(b)db~,
	\label{GHeq}
\end{equation}
with $\beta_0$ to be fixed on physical grounds.
To proceed, it is convenient to discretize the problem by dividing the
normal mass distribution $\rho(M;m,\sigma)$ into $N$ equal
probability and temperature independent sections, each of
which representing a wide energy level, such that
\begin{equation}
	\int_{M_n}^{M_{n+1}}\rho(M;m,\sigma)dM=\frac{1}{N}~.
\end{equation}
This equation is formally solved by invoking the inverse error
function $\erf ^{-1}x$, that is
\begin{equation}
	M_n(\beta)=m(\beta)-\sqrt{2}\sigma(\beta) \,
	\text{erf}^{-1}(1-\frac{2n}{N}) ~,
\end{equation}
for $n=1,...,N-1$.
The normal mass spectrum is depicted in the Fig.{\ref{fig}.
\begin{figure}[h]
	\includegraphics[scale=0.55]{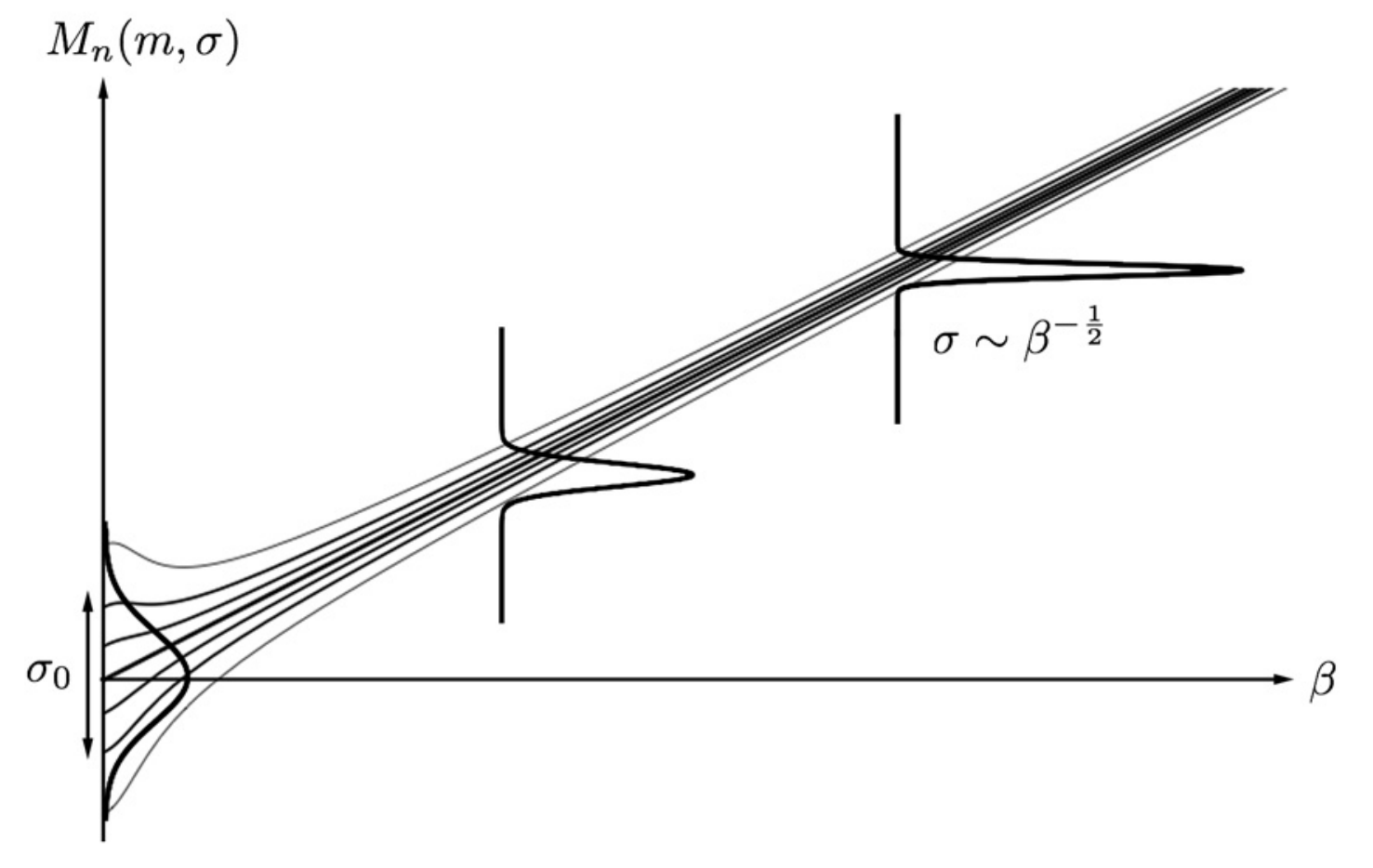}
	\caption{The black hole wave packet mass spectrum
	is plotted as a function of the inverse Hawking temperature
	$\beta$.
	The average mass $m$ is linear in $\beta$.
	The width $\sigma$ peaks at $\sigma_0$ and
	decreases Doppler-like towards the classical limit
	(where the probability of the negative masses
	is practically negligible).}
	\label{fig}
\end{figure}
A straightforward solution of the differential Gibbs-Helmholtz
eq.(\ref{GHeq}), with $M_n (\beta)$ serving as the source term,
reveals the Helmholtz free energy associated with the $n$-th
mass level
\begin{equation}
	\beta F_n=\int_{\beta_0}^{\beta}m(b)db
	-\sqrt{2} \,\text{erf}^{-1}(1-\frac{2n}{N})
	\int_{\beta_0}^{\beta}\sigma(b)db~,
\end{equation}
which we can now substitute into the partition function
$Z(\beta)=\frac{1}{N}\sum _{n=1}^N e^{-\beta  F_n(\beta)}$.
Let $N\rightarrow \infty$, define a continuous integration
variable $\xi=\frac{n}{N}$, and use the integration formula
$\int_0^1 e^{\sqrt{2} \lambda \hspace{1pt} 
\text{erf}^{-1}(1-2 \xi)}\, d\xi=e^{\frac{1}{2}\lambda ^2}$
to arrive at
\begin{equation}
	Z(\beta )=\exp\left(-\int_{\beta _0}^{\beta} m(b)
	\, db+\frac{1}{2} 
	\left(\int_{\beta _0}^{\beta } \sigma (b) \,
	db\right)^2\right)~.
\end{equation}
The entropy $S=\left(1-
\beta \frac{\partial}{\partial{\beta}}\right)\log Z$ can now
be calculated, and turns out to be the sum of two separate
contributions
\begin{equation}
	\boxed{S(\beta)=S_m(\beta)+S_{\sigma}(\beta)}
	\label{S}  
\end{equation}
\begin{subequations}
\begin{eqnarray}
	&&\displaystyle{S_m (\beta)=
	\beta m(\beta)-\int_{\beta_0}^{\beta} m(b) db} ~,  
	\label{Sm}   \\
	&&\displaystyle{S_{\sigma}(\beta)=
	-\beta \sigma(\beta)
	\int_{\beta_0}^{\beta} \sigma(b) db+\frac{1}{2}\left(
	\int_{\beta_0}^{\beta} \sigma(b) db\right)^2} ~.
	\hspace{20pt}
	\label{Ssigma}  
\end{eqnarray}
\end{subequations}
The associated internal energy exhibits a similar structure,
and is given by
\begin{equation}
	U(\beta)=m(\beta)-
	\sigma(\beta)\int_{\beta_0}^{\beta} \sigma(b) db ~,
	\label{U}  
\end{equation}
closing on the first law 
$ S^{\prime}(\beta)=\beta U^{\prime}(\beta)$.

At this stage, $m(\beta)$ and $\sigma(\beta)$ are two
yet unspecified independent functions of $\beta$.
The connection with black hole physics requires some
input beyond the mini super-spacetime model.
This is hereby established by invoking the Bekenstein
area entropy ansatz
\begin{equation}
	\boxed{S=\frac{\langle M^2 \rangle}{2\eta ^2}+c_S}
\end{equation}
quantum mechanically adjusted however by trading the classical
$\langle M \rangle^2=m^2$ for
$\langle M^2 \rangle=m^2+\sigma^2$.
The constant $\eta$ will be recognized as the reduced
Planck mass
\begin{equation}
	\eta=\sqrt{\frac{\hbar c}{8\pi G}}
\end{equation}
as soon as the contact with Hawking temperature is
analytically established, and $c_S$ is a constant to be
determined.
Having the first law for a Gaussian mass
distribution at our disposal, with its compelling 
$m \leftrightarrow\sigma$ split eqs.(\ref{S},\ref{U}),
we can now proceed to calculate the independent functions
$m(\beta)$ and $\sigma(\beta)$.
The corresponding non-linear integral-differential equations
to solve are
\begin{equation}
	S_m (\beta)=\frac{m^2(\beta)}{2\eta^2}+c_m~,~~
	S_{\sigma} (\beta)=\frac{\sigma^2(\beta)}{2\eta^2}
	+c_{\sigma} ~,
\end{equation}
with $c_m+c_{\sigma} =c_S$.

The exact solution of the first equation is straight forward,
and is noticeably $\beta_0,c_S$-independent, namely
\begin{equation}
	m(\beta)=\eta^2\beta ~,
	\label{T}
\end{equation}
reassuring us that the reciprocal Hawking temperature
$\beta$ is proportional, as expected (but non-trivial in
the absence of a sharp horizon), to the necessarily
positive average mass.
The recovery of $m(\beta)$ is a necessary stage preceding
the $\sigma(\beta)$ calculation.
Fixing $\beta_0$ will then determine
$c_m=\frac{1}{2}\eta^2 \beta_0^2$.
The solution of the second equation is somewhat more complicated.
Define
$f(\beta)\equiv \int_{\beta_0}^{\beta}\sigma(b)db$,
so that $\sigma(\beta)=f^{\prime}(\beta)$, and attempt
to solve numerically
\begin{equation}
	f^{\prime}(\beta)=-\eta^2\beta f(\beta)+\eta
	\sqrt{(1+\eta^2\beta^2)f^2(\beta)-2c_{\sigma}}~,
	\label{f}
\end{equation}
subject to $f(\beta_0)=0$.
Before doing so, however, it is crucial to first fix the
$\beta_0$ parameter on physical grounds.

Fowler and Rushbrooke could not give a general rule
for fixing $\beta_0$.
They say "The ambiguity has its counterpart in the use of
the Gibbs Helmholtz equation to derive free energy from
true energy.
One needs to know, for instance, the entropy of the substance
at some one particular temperature".
Under $\beta_0 \rightarrow \beta_0+\delta\beta_0$,
the entropy $S(\beta)$ gets shifted by a $\beta$-dependent
amount.
In other words, the choice of $\beta_0$ is a physical choice.     
And since it cannot be sensitive to $S\rightarrow S+const$,
its roots must be at the
$S^{\prime}(\beta_0)=\beta_0 U^{\prime}(\beta_0)$ level.
The only tenable choice is $\beta_0=0$; it is universal in
the sense that
\begin{equation}
	S(0)=S^{\prime}(0)=U(0)=0 ~,
	\label{universal}
\end{equation}
suggesting (to be implied later)
that $U^{\prime}(0)=\eta^2-\sigma_0^2~$ should vanish
as well. 
In fact, there is even a simpler argument to support the
$\beta_0=0$ choice.
Hawking temperature eq.(\ref{T}) tells us that
choosing $\beta_0$ means choosing a special
average mass $m_0$, but there is no such a special mass.
The accompanying constants take then the values
\begin{equation}
	c_m=0~,~~ c_{\sigma}
	=-\frac{\sigma_0^2}{2\eta^2}=c_S ~.
\end{equation}

Unfortunately, Eq.(\ref{f}) does not admit an exact analytic
solution.
It tells us, however, that
$\sigma(\beta)=f^{\prime}(\beta)$ is a monotonically
decreasing function of $\beta$, solely
parameterized by the maximal width $\sigma_0$.
As far as the small-$\eta\beta$ region is concerned, we
derive the  asymptotic expansion 
\begin{equation}
	\sigma(\beta)=\sigma_0 \left(
	1-\frac{1}{2}\eta^2\beta^2+
	\frac{3}{8}\eta^4\beta^4+... \right)~.
\end{equation}
Even the special case $m=0$, which classically
leads to a flat spacetime, is quantum mechanically
accompanied by a wave packet of non-vanishing width.
Similarly for large-$\eta\beta$, we face
\begin{equation}
	\sigma(\beta) =
	\frac{s\sigma_0}{2\sqrt{\eta\beta}}\left(
	1+\frac{1}{2s^2 \eta\beta}+...\right)~,
\end{equation}
where $s\simeq 0.6185$ has been fixed numerically.
No $\log$-terms \cite{log} at this stage.
The Hawking temperature dependent width
of macro black hole wave packets highly reminds
us (but apparently without any physics in common) of
the Doppler broadening of spectral lines.

$m$ and $\sigma$ have been gradually elevated from
being two independent parameters to two explicit functions
of the Hawking temperature.
Treating $\beta$ as a parameter, one can express
$\sigma(m)$, and proceed to discuss the entropy
and the internal energy.
At the classical limit $m\gg\eta$ there are no surprises,
with the leading Bekenstein-Hawking formulas acquire
only tiny corrections
\begin{equation}
	\begin{array}{rcl}
	 && \displaystyle{S(m)= \frac{m^2}{2\eta^2}
	-\frac{\sigma_0^2}{2\eta^2}
	+\frac{s^2\sigma_0^2}{8\eta m}+...}\vspace{2pt}\\
	 &&  \displaystyle{U(m)=
	 m-\frac{s^2\sigma_0^2}{2\eta}+...}
	\end{array}
\end{equation}
At the quantum regime $m\leq\eta$, on the other hand,
we find ourselves in an unfamiliar territory governed by
\begin{equation}
	\begin{array}{rcl}
	 && \displaystyle{S(m)=
	\left(1- \frac{\sigma_0^2}{\eta^2} \right)
	\frac{m^2}{2\eta^2}
	+\frac{\sigma_0^2 m^4}{2\eta^6}+...}\vspace{2pt}\\
	 &&  \displaystyle{U(m)=
	 \left(1-\frac{\sigma_0^2}{\eta^2}\right)m
	 +\frac{2\sigma_0^2m^3}{3\eta^4}+...}
	\end{array}
\end{equation}
Regarding the value of $\sigma_0$, several
possibilities arise:

\noindent (i) If $\sigma_0=0$, we recover the
Bekenstein-Hawking black hole thermodynamics of 
Schwarzschild spacetime.

\noindent (ii) If $\sigma_0>\eta$, the entropy
function develops a local maximum at $m=0$.
This in turn causes the small-$m$ section of $S(m)$
to be negative, and hence must be rejected on entropy
positivity grounds.

\noindent (iii) If $\sigma_0<\eta$, the entropy
$S(m)$ exhibits an absolute minimum at $m=0$.
The minimal entropy is still proportional to
$S_{BH}= m^2/2\eta^2$, but is suppressed now
by a factor of $1-\sigma_0^2/ \eta^2$.

\noindent (iv) If $\sigma_0=\eta$ (accompanied by
$c_S=-\frac{1}{2}$), the black hole entropy barely
keeps its minimum at $m=0$, and the internal
energy gives up its linear small-$m$ behavior.

The smallest size quantum mechanical black hole
wave packet comes with $m=0$ and $\sigma=\sigma_0$.
We insist on attaching to it a minimal entropy, but do we
have a physical reason which can single out one particular
value for $\sigma_0\leq \eta$?
In fact, we do.
Recall that $U^{\prime}(0)=\eta^2-\sigma_0^2$, so
$\sigma_0=\eta$ can now complete the partial set of
initial conditions eq.(\ref{universal}) by supplementing
the missing piece $U^{\prime}(0)=0$.
Carrying zero entropy, this micro black hole represents
a single degree of freedom, and in this respect can be
regarded elementary.
It is characterized by a finite root mean square mass
$m_{RMS}=\eta$ (consistent with the fact that Compton
wavelength puts a limit on the minimum size of the
region in which a mass can be localized), yet it is
divergently hot, a feature which is supposed to 
play a crucial role at the final stage of black hole
evaporation.

While a classical event horizon is apparently mandatory
for formulating black hole thermodynamics, Bekenstein
entropy will explode and Hawking temperature vanish
if $\hbar$ is switched off.
It is thus relieving to learn that black hole
thermodynamics can be consistently resumed
when $\hbar$ is switched on, causing inevitable horizon
smoothening.
The mass spectrum, which on quantum mechanical
consistency grounds must contain negative masses,
plays here a central role. 
While the average mass sets the recovered Hawking
temperature, a novel temperature dependent width
function contributes to Bekentein entropy.
It peaks at the Planck mass for an elementary quantum
black hole of finite rms size, and decreases
Doppler like towards the classical limit.

\acknowledgments
{We cordially thank BGU president Prof. Rivka Carmi for her
kind support.}

\end{document}